\documentclass[prl,twocolumn,aps,showpacs]{revtex4}
\usepackage{graphicx}% complex graphics
\usepackage{bm}% bold math
\usepackage{amssymb} %math symbols
\begin{document}

\title{Fermi-edge singularity in the vicinity of the resonant scattering condition}

\author{V. V. Mkhitaryan and M. E. Raikh}

\affiliation{ Department of Physics, University of Utah, Salt Lake
City, UT 84112, USA}
\begin{abstract}
Fermi-edge absorption theory predicting the spectrum,
$A(\omega)\propto \omega^{-2\delta_0/\pi+\delta^2_0/\pi^2}$,
relies on the assumption that scattering phase, $\delta_0$, is
frequency-independent. Dependence of $\delta_0$ on $\omega$
becomes crucial near the resonant condition, where the phase
changes abruptly by $\pi$. In this limit, due to finite time spent
by electron on a resonant level, the scattering is {\it dynamic}.
We incorporate this time delay into the theory, solve the Dyson
equation with a modified kernel and find that, near the resonance,
$A(\omega)$ behaves as $\omega^{-3/4} |\ln \omega|$. Resonant
scattering off the core hole takes place in 1D and 2D in the
presence of an empty subband above the Fermi level; then
attraction to hole splits off a resonant level from the bottom of
the empty subband. Fermi-edge absorption in the regime when
resonant level transforms into a Kondo peak is discussed.

\end{abstract}
 \maketitle

{\noindent \it Introduction.} Many-body character of absorption
from the localized level into continuum in the presence of the
Fermi sea \cite{Mahan67} manifests itself via two mechanisms:
scattering of the excited electron from the hole left behind and
adjustment of the Fermi sea to the abrupt switch-on of the hole
potential. Correspondingly, near the threshold, the spectrum,
$A(\omega)=I(\omega)P(\omega)$, is a product \cite{ND} of two
functions
\begin{equation}\label{edge}
I(\omega)=\left(\frac
D\omega\right)^{2\delta_0/\pi}\!\!\!\!,\qquad
P(\omega)=\left(\frac D\omega\right)^{-\delta^2_0/\pi^2}\!\!\!\!,
\end{equation}
where $D$ is the bandwidth; $\delta_0$ is the phase shift for $s$-
scattering.
%from potential $V\delta({\bf r})$ [for which
%scattering occurs only in the $s$-channel] of the hole left
%behind.
Phase $\delta_0$ is related to the hole potential strength $V$ as
$\tan \delta_0=\pi\nu_0V$, where $\nu_0$ is the density of states.
The function $I(\omega)$ is the single-electron-line contribution,
which describes the first mechanism, while $P(\omega)$ accounts
for the shake-up effects.
%%%%%%%%%%%%%%%%%%%
\begin{figure}[b]
\centerline{\includegraphics[width=85mm,angle=0,clip]{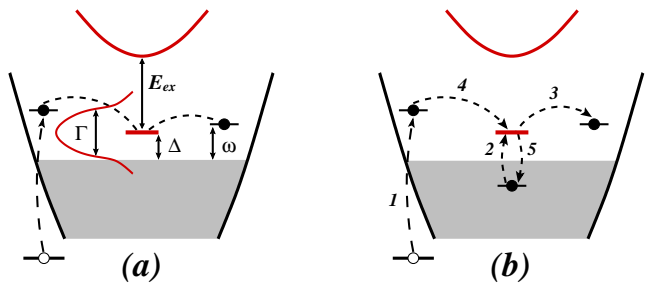}}
\caption{(Color online) (a): In the vicinity of a topological
transition, the core hole potential creates bound state (resonant
level) with energy, $E_{ex}$, measured from the bottom of the
upper subband; its lifetime, $\Gamma$, is due to a coupling to the
lower subband. For a photoelectron with energy, $\omega\ll\Gamma$
above the Fermi level, dynamical character of scattering from
resonant level is crucial. (b): Contribution to the absorption
amplitude which is strongly affected by resonant character of
scattering: two scattering acts involved, ($2+3$) and ($4+5$),
have opposite signs of amplitudes, Eq. (\ref{scatamp}). }
\label{twoband}
\end{figure}
%%%%%%%%%%%%%%%%%%

Derivation of Eq. (\ref{edge}) by Nozieres and De Dominicis (ND)
in Ref. \cite{ND} strongly relies on the assumption that
$\delta_0$ is frequency-independent. This assumption implies that
the scattering is {\it instantaneous}. In the present paper we
focus on the situation when this assumption is violated. This
situation realizes when the Fermi level is close to an empty upper
subband, as shown in Fig. \ref{twoband}. In such an arrangement it
is important that even a {\it deep} core-hole creates a localized
level a distance $E_{ex}$ below the bottom of the upper subband,
see Fig. \ref{twoband}. Formation of a localized level happens in
low dimensions (2D and 1D). Experimental  and theoretical studies
of the arrangement in which Fermi level is near the bottom of
empty subband have been reported in \cite{Skolnick91, Chen, Melin,
Kopelevich} and \cite{Matveev, Starykh, Ablyazov, Alicea, Muller,
Shahbazyan, Balents2000}, respectively.

Due to degeneracy with continuum of lower band, the level $E_{ex}$
acquires a finite width, $\Gamma$. Then the process underlying the
Fermi-edge singularity is not scattering from {\it bare} core-hole
potential but rather resonant scattering on the quasi-local level.
Correspondingly, the scattering pase is given by
\begin{equation}
\label{scatphase}\delta(\omega)=\text{arctan}\frac\Gamma {\omega
-\Delta},
\end{equation}
where $\Delta$ is the energy distance from localized level to the
Fermi level, $E_{\scriptscriptstyle F}$. In the limit
$\Delta\rightarrow0$, $\omega$-dependence of $\delta$ is strong.
Indication that in this limit the ND theory is inapplicable
follows from the fact that at $\omega=\Delta$, the phase abruptly
changes by $\pi$. This, in turn, leads to a paradoxical conclusion
that, as Fermi level is swept through the bound state, $I(\omega)$
changes {\it abruptly} from $I(\omega)\propto 1/\omega$ to
$I(\omega)\propto \omega$. By contrast, the shake-up term
$P(\omega)\propto \omega^{1/4}$ {\it does not} experience a jump.

\begin{figure}[t]
\centerline{\includegraphics[width=80mm,angle=0,clip]{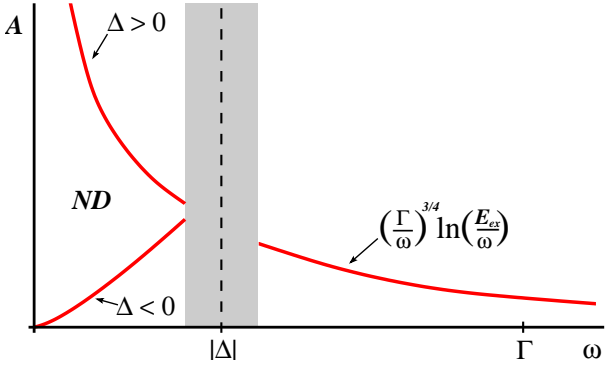}}
\caption{(Color online) Absorption spectrum, $A(\omega)$, in the
situation when the Fermi level is close to resonant level
($|\Delta|\ll\Gamma$) is shown schematically. For
$\omega<|\Delta|$ the $\omega$-dependence, Eq. (\ref{withdelta}),
is governed by ND theory. In the "dynamic scattering" domain
$|\Delta|<\omega<\Gamma$ the spectrum is
$A(\omega)=(\Gamma/\omega)^{3/4}\ln(E_{ex}/\omega)\ln(E_{ex}/\Gamma)$
for {\it any sign} of $\Delta$.} \label{graph}
\end{figure}

As a resolution of this paradox, in the present paper we show that
abrupt change of $I(\omega)$ {\it does} take place in the
frequency interval, $\omega\sim|\Delta|$, which gets {\it
progressively narrow} as $\Delta$ goes to $0$. We also show that
outside this interval, where the ND theory does not apply, a new,
{\it dynamical resonant-scattering} regime governs the absorption
\cite{footnote}. We show that in this regime, instead of Eq.
(\ref{edge}), $I(\omega)$ is given by
\begin{equation}
\label{result} I(\omega){\Big
|}_{|\Delta|<\omega<\Gamma}=\frac{\nu_0}{2\pi}\frac\Gamma{\omega}
\ln\left(\frac{E_{ex}}\omega\right),
%\ln\left(\frac{E_{ex}}\Gamma\right),
\end{equation}
{\it independently} of the sign of $\Delta$. Our findings are
illustrated in Fig. \ref{graph}.

\begin{figure}[t]
\centerline{\includegraphics[width=75mm,angle=0,clip]{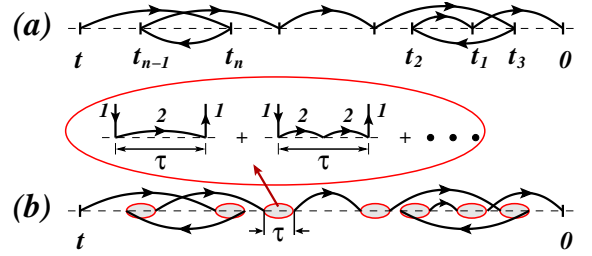}}
\caption{(Color online) (a): Typical contribution, Eq.
(\ref{nth}), to the transient Green function in ND theory. {\it
Instantaneous} scatterings take place at time moments, $t_1$,
$t_2$,...,$t_n$. (b): In the presence of a resonant level, each
scattering act involves a "visit" to the resonant level, and takes
a finite time, $\tau$. Blowup illustrates a ladder of scatterings
off the core-hole in the upper subband, $2$, which lead to
resonant scattering in the lower subband, $1$.} \label{diagrams}
\end{figure}

{\noindent \it Dynamic-scattering regime.} Within the ND theory,
the transient Green function,
$\varphi(\tau,\tau^\prime|t,t^\prime)$, satisfies the Dyson
equation
\begin{equation}
\label{NDinteq} \varphi(\tau,\tau^\prime|t,t^\prime)=
G(\tau-\tau^\prime)+\!\!\int_t^{t^\prime}\!\!\!\!
d\tilde{\tau}K(\tau,\tilde{\tau})
\varphi(\tilde{\tau},\tau^\prime|t,t^\prime),
\end{equation}
where $G(t)=-\nu_0/t$ is the bare electron Green function, and the
kernel is defined as $K(\tau,\tilde{\tau})= V
G(\tau-\tilde{\tau})$. The single-electron-line contribution to
$A(\omega)$ is expressed in terms of the transient Green function
as
\begin{equation} \label{viatgf}
I(\omega)=-\frac1\pi\,\text{Im}\!\int_{-\infty}^0\!\!dt
\,e^{-i\omega t}\varphi(0^-,t^+|t,0).
\end{equation}
Delicate character of Fermi-edge absorption can be inferred from
the structure  of $n$-th term,
\begin{equation}
\label{nth} V^n\!\!\int_t^0\!\!\!dt_1\cdots\int_t^0\!\!\!dt_n
\,G(-t_1)G(t_1-t_2) \cdots G(t_n-t),
\end{equation}
of perturbative expansion of $\varphi(0^-,t^+|t,0)$ in powers of
$V$, which follows from Eq. (\ref{NDinteq}). It appears that
contributions of different orderings of times $t_i$, Fig.
\ref{diagrams}a, at which electron is scattered, cancel each other
up to $1/n!$.
%%%%%%%%%%%%%%%%%%%
%%%%%%%%%%%%%%%%%%

Frequency dependence of $\delta$ leads to a time delay, $\tau$, in
the scattering processes, as illustrated in Fig. \ref{diagrams}b.
Each delay ranges from $E_{ex}^{-1}$ to $|\Delta|^{-1}$. As a
result, in the limit $\Delta\rightarrow0$, the above cancellation
is {\it completely destroyed.} On the quantitative level,
dynamical character of the resonant scattering alters the kernel
of Eq. (\ref{NDinteq}).
%%%%%%%%%%%%%%%%%%%
%%%%%%%%%%%%%%%%%%

{\noindent \it Resonant-scattering kernel.} In the presence of the
empty upper subband, the constant interaction $V$ for electrons of
the Fermi sea acquires a frequency-dependence of the form
\begin{equation} \label{effint} \tilde{V}(\omega)= V_{12}
\frac{G_2(\omega)}{1-V_{22}G_2(\omega)} V_{21}.
\end{equation}
Here $V_{12}$ and $V_{22}$ are inter- and intra-subband matrix
elements of the hole potential, respectively; $G_2$ is the Green
function of the upper subband,
\begin{equation} \label{G2}
G_2(\omega)=\sum_q \frac1 {\omega-\epsilon_{2q}+i\eta},
\end{equation}
and $\epsilon_{2q}=\hbar^2q^2/2m$ is the spectrum near the bottom.
The form Eq. (\ref{effint}) is a result of summation of
ladder-type diagrams, see a blowup in Fig. \ref{diagrams}. Bound
state emerges as a pole in Eq. (\ref{effint}),
\begin{equation} \label{Eex}
E_{ex}=\frac{2\pi^2m}{\hbar^2}|V_{22}|^2\,\, \text{(1D)},\quad
E_{ex}=De^{-\frac{2\pi\hbar^2}{m|V_{22}|}}\,\, \text{(2D)}.
\end{equation}
Expanding near the pole, we simplify $\tilde{V}$ to the form
\begin{equation}
\label{scatamp}
\tilde{V}(\omega)=\left(\frac\Gamma{\pi\nu_0}\right)\frac 1{\omega
-\Delta +i\eta},
\end{equation}
where width, $\Gamma$, is given by
\begin{equation} \label{Eex}
\Gamma=\frac{2\pi\nu_0 E_{ex}}{|V_{22}|}|V_{12}|^2\,\,
\text{(1D)},\,\,\,
\Gamma=\frac{2\pi^2\hbar^2\nu_0E_{ex}}{m|V_{22}|^2}|V_{12}|^2\,\,
\text{(2D)}.
\end{equation}
Effective interaction Eq. (\ref{scatamp}) corresponds to {\it
repulsion} for $\omega>\Delta$ and to {\it attraction} for
$\omega<\Delta$. In the resonant-scattering regime, the kernel in
Eq. (\ref{NDinteq}) instead of the simple product $V
G(\tau-\tilde{\tau})$ becomes a convolution
\begin{equation}
\label{reskern} K_{res}(\tau,\tau^\prime)=
\int\limits_t^0\!\!d\tau^\ast
G(\tau-\tau^\ast)\!\int\limits_{-\infty}^\infty\!\!\frac{d\omega}{2\pi}
e^{-i\omega(\tau^\ast-\tau^\prime)}\tilde{V}(\omega).
\end{equation}
Integration over $\omega$  can be carried out explicitly, yielding
\begin{equation}
\label{expreskern} K_{res}(\tau,\tau^\prime)=i\frac\Gamma\pi
\int_{t}^0d\tau^\ast
\frac{\theta(\tau^\ast-\tau^\prime)}{\tau-\tau^\ast}
e^{-i\Delta(\tau^\ast-\tau^\prime)}.
%\ln\!\left(\frac{\tau}{|\tau-\tau^\prime|}\right).
\end{equation}
For most interesting case, $\Delta=0$, the expression for the
kernel assumes the form
\begin{equation}
\label{expkern} K_{res}(\tau,\tau^\prime)=-i\frac\Gamma\pi
\ln{\Big |}\frac{\tau}{\tau-\tau^\prime}{\Big |}.
\end{equation}
We see that $\ln|\tau-\tau^\prime|$ in $K_{res}$ emerges in the
place of $1/(\tau-\tau^\prime)$ in the ND kernel. This is a
consequence of {\it opposite signs} of interaction with resonant
level for electrons and for holes.

{\noindent \it Derivation of Eq. (\ref{result}).}  Substituting
Eq. (\ref{expkern}) into Eq. (\ref{NDinteq}) and performing the
rescaling
\begin{equation}
\label{xt} \varphi(\tau,\tau^\prime|t,0)= \frac{\nu_0}t\,
\phi\!\left( \frac\tau t,\,\frac{\tau^\prime} t\right),
\end{equation}
we arrive to the following dimensionless equation
\begin{equation}
\label{dlessieq} \phi(x,y)= -\frac1{x-y}+i\gamma\!\!\int_0^1\!\!dz
\ln\left(\!\frac{x}{|x-z|}\!\right) \phi(z,y),
\end{equation}
where $\gamma=\frac\Gamma\pi t$. Note that the kernel Eq.
(\ref{dlessieq}) is much less singular than the ND kernel. As a
result, the power-law transient factor in the ND solution,
$[(1-x)y/(1-y)x]^{\delta/\pi}$, which is singular for
$x\rightarrow0$, $y\rightarrow 1$, does not emerge. This allows us
to search for solution of Eq. (\ref{dlessieq}) as a linear
combination,
\begin{equation} \label{expand}
\phi(x,y)=\sum_nc_n(y,\gamma)u_n(x),
\end{equation}
of the eigenfunctions, $u_n(x)$, of the Hermitian integral
operator
\begin{equation}
\label{intop}
\hat{R}\,\{u_n(x)\}=\int_0^1dz\ln|x-z|u_n(z)=\lambda_nu_n(x),
\end{equation}
where $\lambda_n$ are the eigenvalues. The kernel in Eq.
(\ref{intop}) is symmetric with respect to $x=1/2$.
Correspondingly, $u_n(x)$, $n=0,1,2,\cdots$ are even (for $n$
even) and odd (for $n$ odd) functions of $(x-\frac12)$.
Substitution of Eq. (\ref{expand}) into both sides of Eq.
(\ref{dlessieq}) yields the following expression for coefficients
of expansion,
\begin{equation} \label{cm}
c_m(y,\gamma)=\frac{a_m(y)+b_md(y,\gamma)} {1+i\gamma\lambda_m},
\end{equation}
where
\begin{eqnarray}
\label{not1} &&a_m(y)=\!\int_0^1\!\!dz\,\frac{u_m(z)}
{y-z}=\lambda_mu_m^\prime(y),\\
&&b_m=\!\int_0^1\!\!dz\,u_m(z)\ln z=\lambda_m u_m(0). \label{not2}
\end{eqnarray}
In the second identities of Eqs. (\ref{not1}), (\ref{not2}) we
used the properties of operator $\hat{R}$.

The kernel, $\ln x-\ln|x-y|$, of Eq. (\ref{dlessieq}), in addition
to the difference term, contains a $y$- independent term. As a
result, $b_m$ in the expression for $c_m(y,\gamma)$ enters with
coefficient
\begin{equation} \label{not3}
 d(y,\gamma)= i\gamma\int_0^1dz\,
\phi(z,y),
\end{equation}
which is {\it the same} for all $m$, so that Eq. (\ref{not3}) can
be viewed as a self-consistency condition.

Our key observation which will be justified later is that, the
relevant values of $m$ in the expansion Eq. (\ref{expand}) are
$\gg1$. For such $m$, it can be shown that eigenvalues behave as
$\lambda_m\approx -1/m$. Concerning the eigenfunctions $u_m(x)$,
they assume constant value near the boundaries, $u_m(0)=\pm
u_m(1)$, while the derivatives, $u_m^\prime(x)$ diverge
logarithmically at the boundaries. Outside of small ($\sim 1/m$)
intervals from the boundaries, they behave as $\sin(2\pi
x/\lambda_m)$ and $\cos(2\pi x/\lambda_m)$. All these large- $m$
properties can be established upon integrating by parts in Eq.
(\ref{intop}). Then the remaining integral will be determined by a
narrow domain, $|x-z|\sim1/m$. The same simplification allows us
to obtain a concise expression for $d(y,\gamma)$, namely,
\begin{equation}
\label{df} d(y,\gamma)\simeq-i\gamma\ln(1-y),
\end{equation}
and subsequently the solution for $\phi(x,y)$ in the form
\begin{equation}
\label{sinsol} \phi(x,y)\simeq -\ln(1-y)
\sum_n\frac{[u_n(1)+i\gamma\lambda_nu_n(0)]u_n(x)}
{1+i\gamma\lambda_n}.
\end{equation}
In calculating $I(\omega)$ from Eq. (\ref{sinsol}) the
combination, $\sum_nu_n(1)u_n(x)$, appears. This combination is
equal to $\delta(1-x)$ and does not contribute to absorption,
which is determined by $x\rightarrow 0$ and $y\rightarrow 1$. In
ND problem, divergences in the time domain are cut off by the
inverse bandwidth, $iD^{-1}$. In our case, the logarithmic
divergence in Eq. (\ref{sinsol}) is terminated at
$(1-y)=i(E_{ex}t)^{-1}$. Correspondingly, the minimal $x$ in Eq.
(\ref{sinsol}) is, in fact, $x_{min}=i(E_{ex}t)^{-1}$. Taking this
into account and using Eqs. (\ref{viatgf}) and (\ref{xt}), we
express $I(\omega)$ in the integral form
\begin{equation}
\label{ans1} I(\omega)\simeq \frac{2\nu_0\Gamma}\pi
\text{Im}\!\int^{\infty}_0 \!\!dt\, e^{i\omega t}\,
S(t)\ln(iE_{ex}t),
\end{equation}
where the sum,
\begin{equation}\label{sum}
S(t)=\sum_{ \text{odd}\,\,
m}\frac{\lambda_{m}u_{m}(0)u_m(x_{min})} {\lambda_{m} \Gamma
t+i\pi},
\end{equation}
is performed over only odd $m$, for which $u_m(1)=-u_m(0)$.
Summation over $m$ requires large-$m$ values of $u_m(0)$. In fact,
these values saturate with increasing $m$. This follows from the
identity
\begin{equation}
\label{logexp} \sum_{\text{odd}\,\, m}
\lambda_mu_m(0)u_m(x)=\frac12\ln\left(\frac x{1-x}\right).
\end{equation}
The left-hand side has a logarithmic divergence at
$x\rightarrow0$. In the right-hand side, with $\lambda_m\approx
-1/m$, logarithmic divergence of the sum is achieved with
$u_m(x)\approx u_m(0)\approx 1$ for $m<x^{-1}$. Eq. (\ref{logexp})
is a direct consequence of the identity, $\ln x=\sum_m\lambda_m
u_m(0)u_m(x)$. The sum over odd $m$ appears upon writing this
identity for $x$ and $(1-x)$ and taking their difference.
Comparing Eqs. (\ref{sum}) and (\ref{logexp}) we conclude that
\begin{equation}\label{sumans}
S(t)\approx\frac14+\frac1{2\pi i}\ln(i\Gamma
tx_{min})=\frac14+\frac1{2\pi i}\ln
\left(\frac{E_{ex}}{\Gamma}\right).
\end{equation}
Imaginary part of $S(t)$ is determined by the upper cut-off, while
$\text{Re}S(t)$ comes from $m\sim \Gamma t$; relevant values of
$m$ are large, as we assumed above. Factor $\Gamma t$ in the
argument of the logarithm comes from the terms in sum Eq.
(\ref{sum}) with $m\lesssim\Gamma t$.

With $S(t)$ being time-independent for $t\gg\Gamma^{-1}$,
integration in Eq. (\ref{ans1}) recovers our result Eq.
(\ref{result}). Physical meaning of the enhancement factor,
$\Gamma/\omega$, in Eq. (\ref{result}) is the number of times
electron virtually visits the resonant level and returns back to
the lower subband during the time $1/\omega$.

\noindent{\em Concluding remarks.} ({\it i}) Our main result, Eq.
(\ref{result}), was obtained by setting $\Delta=0$ in the kernel
Eq. (\ref{expreskern}). Finite-$\Delta$ correction to $A(\omega)$
is small as $(\Delta/\omega)^2$. Let us briefly discuss the
opposite limit, $\omega\ll|\Delta|\ll\Gamma$. In this limit, the
characteristic time $t$ in Fig. \ref{diagrams} is
$\gg|\Delta|^{-1}$. Then, our above consideration for $\Delta=0$
applies only in the time domain $(-|\Delta|^{-1},0)$. In the
remaining domain $(t,-|\Delta|^{-1})$ characteristic time
intervals $t_i-t_{i+1}$ in Fig.~\ref{diagrams} are much bigger
than $|\Delta|^{-1}$. This makes scattering "instantaneous", and
ND theory applicable in the domain $(t,-|\Delta|^{-1})$;
corresponding time-independent scattering amplitude and effective
bandwidth are $\Gamma/(\pi\nu_0\Delta)$ and $\Delta$,
respectively. Separation of time scales leads to the absorption
coefficient in the form of a product of
$(\omega/|\Delta|)^{-2\text{arctan}(\Gamma/\Delta)/\pi}\approx
(\omega/|\Delta|)^{-\text{sign}\Delta}$ (from long times), and
$\Gamma/|\Delta|$ (from short times, $|t_i|< |\Delta|^{-1}$). With
shake-up included, the overall result for absorption coefficient,
\begin{equation}\label{withdelta}
A(\omega){\Big |}_{\Delta>0}\!\!\propto \frac\Gamma\omega
\left(\frac\omega\Gamma \right)^{\frac14}\!\!,\qquad
A(\omega){\Big |}_{\Delta<0}\!\!\propto
\frac{\omega\Gamma}{\Delta^2} \left(\frac\omega\Gamma
\right)^{\frac14}\!\!\! ,
\end{equation}
is illustrated in Fig. \ref{graph}. At $\omega\sim\Delta$, Eq.
(\ref{withdelta}) matches Eq. (\ref{result}) with logarithmic
accuracy.

({\it ii}) Ellipses in Fig. \ref{diagrams} can overlap; this
corresponds to double-occupancy of the resonant level, which our
theory does not prohibit. This implies that the Hubbard repulsion,
$U$, is smaller than $\Gamma$. Since $U$ is inversely proportional
to the radius of the level wave function, this condition can be
met.

Even more interesting is the situation when resonant level, which
is split off the upper band by the core-hole, is well below the
Fermi level ($|\Delta|\gg \Gamma$), and $U$ is not small. Then the
level gets occupied, and after time $\sim T_K^{-1}\sim
\exp[\pi\Delta/2\Gamma]$ a Kondo peak of width $\sim T_K$ is
formed at the Fermi level \cite{Pustilnik}. Photoexcited electron
will experience dynamical resonant scattering from this peak. One
can argue that our result for dynamical resonant scattering regime
will apply in this situation, with replacement,
$(\Gamma/\omega)\rightarrow(T_K/\omega)$. However, we cannot make
this analogy rigorous because the Kondo scattering of the state
$\omega$ leads to finite lifetime $\sim T_K/\omega^2$.

Another interesting possibility is then the core-hole possesses
spin \cite{Our}. Then a Kondo peak at the Fermi level will form
even without the upper subband, so that the scattering off the
core-hole will become a Kondo-scattering with $\omega$-dependent
scattering phase.

({\it iii}) We stated above that, unlike  the single-electron-line
contribution, $I(\omega)$, the shake-up factor, $P(\omega)=
(\omega/\Gamma)^{\,\text{arctan}^2(\Gamma/\Delta)/\pi^2}\approx
(\omega/\Gamma)^{1/4}$, is not sensitive to the sign of the
detuning, $\Delta$. To prove this statement rigorously, one has to
extend evaluation of the many-body overlap integral Ref.
\cite{Hamann} to the case of resonant scattering with
frequency-dependent scattering amplitude $\tilde{V}(\omega)$. The
modified overlap integral reads
\begin{eqnarray}
\label{overlapsol} && -\frac1{(2\pi)^2}\int_{0}^1\!\!\!d\lambda\!
\int_{-\infty}^\infty\!\!\!d\Omega \frac{
\tilde{V}(\Omega)\,G(\Omega)}{1-\lambda
\tilde{V}(\Omega)\,G(\Omega)} \nonumber \\
&&\,\nonumber \\
%&&\hspace{2cm}
&&\times\int_{-\infty}^\infty\!\!\!d\Omega^\prime\,
\frac{\ln[1-\lambda \tilde{V}(\Omega^\prime)\,G(\Omega^\prime)]}
{(\Omega^\prime-\Omega+i\eta)^2}.
\end{eqnarray}
For $\tilde{V}(\Omega)=\text{const}(\Omega)$, the integral Eq.
(\ref{overlapsol}) comes from small arguments, namely
$\Omega\rightarrow0^+$ and  $\Omega^\prime\rightarrow0^-$. It
diverges logarithmically with coefficient $-(\delta_0/\pi)^2$,
where $\delta_0= -\text{Arg}[1-VG_r(0)]$, and $G_r$ is the
retarded Green function. Using the analytical properties of
resonant scattering amplitude, Eq. (\ref{scatamp}), one can check
that the above logarithmic divergence persists when $\tilde{V}$ is
frequency-dependent. Independence of the sign of $\Delta$ follows
from the fact that coefficient in front of logarithmically
divergent part is $\frac1{\pi^2}\,
\delta_\Delta(0)\delta_{-\Delta}(0)$, where
$\delta_\Delta(\omega)=\tan^{-1}[\Gamma/(\omega-\Delta)]$ is the
resonant-scattering phase.

({\it iv}) We neglected a contribution to absorption from a direct
transition core-level $\rightarrow$ $E_{ex}$. Since the width of
$E_{ex}$ is $\sim\Gamma$, this contribution is non-singular at
small $\omega$.

({\it v}) Fermi edge physics manifests itself also in resonant
tunneling via an impurity \cite{LarMatLev}. However, the regime
$\Delta\ll\Gamma$, considered in the present paper, can not be
realized in this setting since at long times impurity gets
reoccupied.

\noindent{\em Acknowledgments.} We are grateful to T. V.
Shahbazyan for very helpful discusssions.


\begin{references}
\bibitem{footnote} Dynamic and non-equilibrium aspects of conventional
Fermi-edge singularity have been considered in Refs.
\cite{dynamic} and \cite{noneq}, respectively.
\end{references}
\end{document}